%% file: main.tex
\documentclass[%
reprint,
superscriptaddress,
amsmath,amssymb,
aps,
prl,
]{revtex4-2}

\usepackage{graphicx}
\usepackage{dcolumn}
\usepackage{bm}
\usepackage[hypertexnames=false]{hyperref}

\usepackage[T1]{fontenc}
\usepackage{newtxtext,newtxmath}
\usepackage{braket}
\usepackage{comment}
\usepackage{color}

\begin{document}

\renewcommand{\figurename}{Fig.}

\title{Universal constraint on nonlinear population dynamics}

\author{Kyosuke Adachi}
\affiliation{Nonequilibrium Physics of Living Matter RIKEN Hakubi Research Team, RIKEN Center for Biosystems Dynamics Research (BDR), 2-2-3 Minatojima-minamimachi, Chuo-ku, Kobe 650-0047, Japan}
\affiliation{RIKEN Interdisciplinary Theoretical and Mathematical Sciences Program (iTHEMS), 2-1 Hirosawa, Wako 351-0198, Japan}

\author{Ryosuke Iritani}
\affiliation{RIKEN Interdisciplinary Theoretical and Mathematical Sciences Program (iTHEMS), 2-1 Hirosawa, Wako 351-0198, Japan}
\affiliation{Department of Biological Sciences, Graduate School of Science, University of Tokyo, 7-3-1 Hongo, Bunkyo-ku, Tokyo 113-0033, Japan}
\author{Ryusuke Hamazaki}
\affiliation{Nonequilibrium Quantum Statistical Mechanics RIKEN Hakubi Research Team, RIKEN Cluster for Pioneering Research (CPR), 2-1 Hirosawa, Wako 351-0198, Japan}
\affiliation{RIKEN Interdisciplinary Theoretical and Mathematical Sciences Program (iTHEMS), 2-1 Hirosawa, Wako 351-0198, Japan}

\date{\today}

\begin{abstract}
Ecological and evolutionary processes show various population dynamics depending on internal interactions and environmental changes.
While crucial in predicting biological processes, discovering general relations for such nonlinear dynamics has remained a challenge.
Here, we derive a universal information-theoretical constraint on a broad class of nonlinear dynamical systems represented as population dynamics.
The constraint is interpreted as a generalization of Fisher's fundamental theorem of natural selection.
Furthermore, the constraint indicates nontrivial bounds for the speed of critical relaxation around bifurcation points, which we argue are universally determined only by the type of bifurcation.
Our theory is verified for an evolutionary model and an epidemiological model, which exhibit the transcritical bifurcation, as well as for an ecological model, which undergoes limit-cycle oscillation.
This work paves a way to predict biological dynamics in light of information theory, by providing fundamental relations in nonequilibrium statistical mechanics of nonlinear systems.
\end{abstract}

\maketitle


{\noindent \textbf{Introduction}}

Nonlinear dynamics appears in a variety of fields, including classical mechanics, chemical reaction systems, and population biology, to name a few~\cite{Strogatz2018}.
Nonlinearity can trigger complex temporal and spatial patterns and even chaotic behaviors, making it challenging to find universal relations within the properties of dynamics.
In particular, slight perturbations in external parameters can result in qualitative changes in the dynamical property through a bifurcation such as the Hopf bifurcation, where self-sustained oscillation emerges.
It is of pivotal importance to explore universal relations shared by a broad class of dynamical phenomena with nonlinearity.

Ecological and evolutionary processes often exhibit nonlinear population dynamics~\cite{Levine2017,Hastings2018} such as temporal oscillation in population sizes and irreversible extinction of certain species~\cite{Hofbauer1998}.
Typical biological systems consist of identifiable units such as genotypes and species (called ``types’’ in this paper), and intra-type and inter-type interactions cause nonlinear dynamics~\cite{Hofbauer1998,Levine2017}.
Besides interactions, type-dependent growth rates determined by natural selection lead to nonlinear dynamics of the proportions of each type.
In evolutionary theory, Fisher's fundamental theorem of natural selection~\cite{Li1967,Edwards1994} establishes a simple relation between the variance of the growth rate and the temporal increase in the average growth rate.
Though the theorem has been extended to ecological models~\cite{Baez2021}, mutation processes have been outside the scope of the theorem.

Bifurcations and associated critical dynamics play significant roles in biological processes~\cite{Munoz2018}.
In ecological~\cite{Veraart2012,Dai2012} and epidemiological~\cite{Drake2019} systems, critical slowing down around bifurcation points has been discussed as an early warning signal for catastrophic shifts.
In evolutionary systems, bifurcation points can appear as critical mutation rates beyond which heredity does not persist~\cite{Bull2005,Sole2021}, and the self-organized criticality has also been discussed as a possible mechanism of mass extinction of species~\cite{Sneppen1995}.
Since such critical dynamics reflects instabilities behind nonlinear systems~\cite{Scheffer2015}, fundamental relations near bifurcation points are crucial in predicting dramatic changes in ecological and evolutionary processes.

We here derive a general constraint on nonlinear population dynamics by extending the formulation developed for stochastic processes~\cite{Ito2020,Nicholson2020} to nonlinear dynamical systems.
In particular, Fisher's fundamental theorem of natural selection is a special case of the constraint.
As a unique consequence of the constraint, we show that the critical scaling exponents of speeds near the bifurcation point should have nontrivial bounds that are universally determined by the type of bifurcation.
We verify our theory for an evolutionary model with mutation and the SIR model with birth and death, which show the transcritical bifurcation, as well as for the competitive Lotka-Volterra model, which undergoes limit-cycle oscillation.
\\

\begin{figure*}[t]
\includegraphics[scale=0.85]{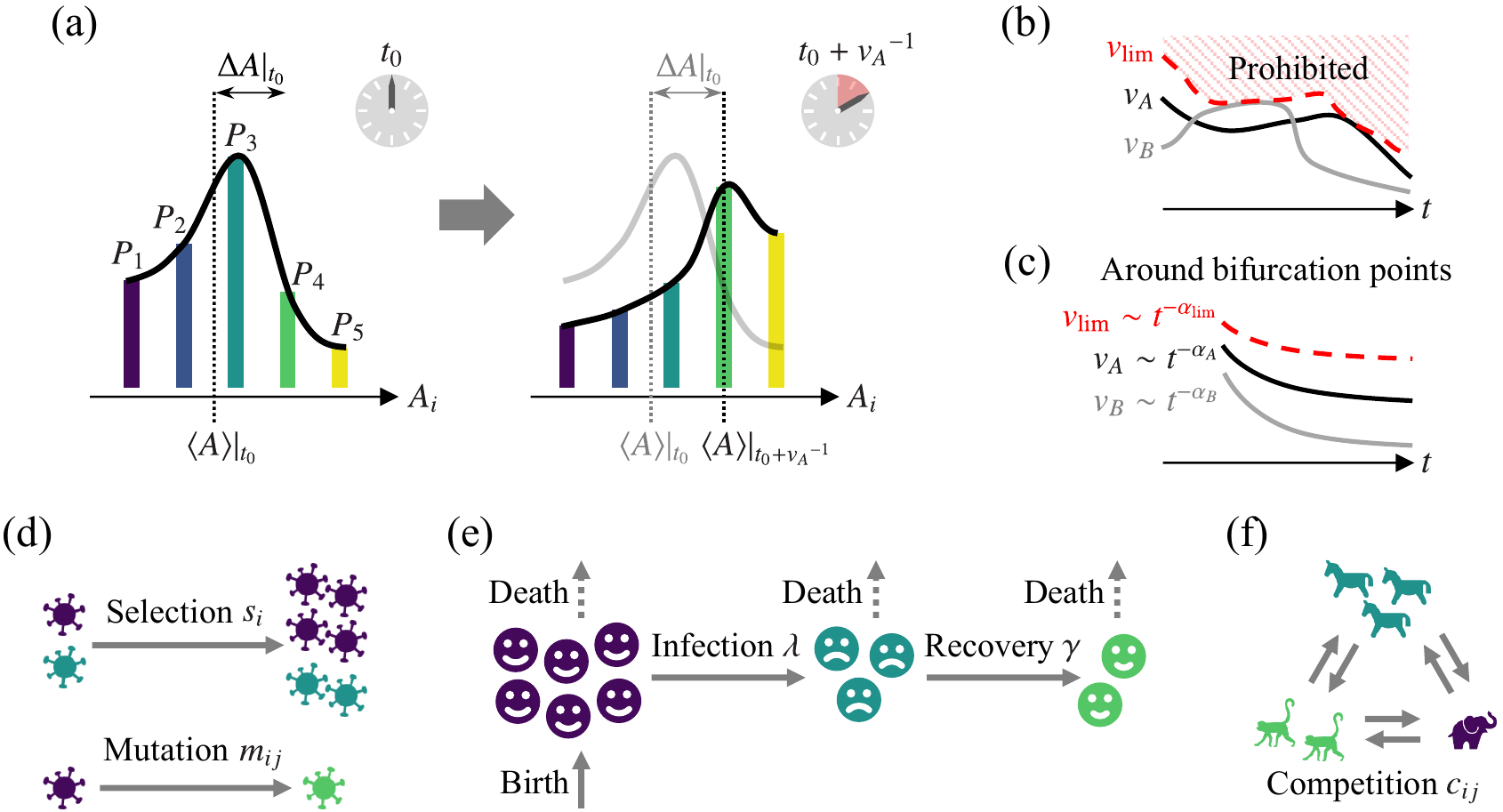}
\caption{\textbf{Speed-limit inequality in ecological and evolutionary dynamics.}
(a) Inverse of the speed, ${v_A}^{-1}$, represents the time required for the instantaneous average $\braket{A}$ to change by the instantaneous standard deviation $\Delta A$.
In (a), we assume $A_1 < A_2 < A_3 < A_4 < A_5$ without loss of generality.
(b) For any quantity $A$ and at any time, speeds faster than $v_\mathrm{lim}$ are prohibited.
(c) Around bifurcation points, the speed and speed limit show power-law decays as $v_A \sim t^{-\alpha_A}$ and $v_\mathrm{lim} \sim t^{-\alpha_\mathrm{lim}}$ with a constraint $\alpha_A \geq \alpha_\mathrm{lim}$ for any $A$, where $\alpha_\mathrm{lim}$ is universally determined by the bifurcation type.
In this study, we mainly consider three models: (d) the evolutionary model with natural selection and mutation, (e) the SIR model, and (f) the competitive Lotka-Volterra model.}
\label{Fig:Overview}
\end{figure*}


{\noindent \textbf{Results}}

{\noindent \textbf{Constraint on general population dynamics.}}
We consider a general population dynamics described by
\begin{equation}
    \partial_t N_i = F_i(N_1, ..., N_L),
    \label{Eq:GenericPopDynamics}
\end{equation}
where $i$ is the label for each type, $L$ is the total number of types, and $N_i (t)$ is the density of type $i$ at time $t$.
If there are interactions between types, $F_i (N_1, ..., N_L)$ is generally a nonlinear function.
Defining the proportion $P := \{ P_i \}_{i = 1}^L := \{ N_i / N_\mathrm{tot} \}_{i = 1}^L$ with the total population density $N_\mathrm{tot} := \sum_{i = 1}^L N_i$, we obtain equations for $P_i$ and $N_\mathrm{tot}$ as
\begin{equation}
    \partial_t P_i = \frac{F_i (N_\mathrm{tot} P_1, ..., N_\mathrm{tot} P_L)}{N_\mathrm{tot}} - P_i \sum_{j = 1}^L \frac{F_j (N_\mathrm{tot} P_1, ..., N_\mathrm{tot} P_L)}{N_\mathrm{tot}}
    \label{Eq:GenericFreqDynamics}
\end{equation}
and $\partial_t N_\mathrm{tot} = \sum_{i = 1}^L F_i (N_\mathrm{tot} P_1, ..., N_\mathrm{tot} P_L)$, respectively.
Even if $F_i (N_1, ..., N_L)$ is a linear function for all $i$, Eq.~\eqref{Eq:GenericFreqDynamics} can be a nonlinear equation, and bifurcations can occur as we discuss later.

Applying the Cauchy-Schwarz inequality to the Price equation~\cite{Price1972a,Frank2020}, which is derived from the conservation of the total proportion ($\sum_{i = 1}^L P_i = 1$), we obtain the speed-limit inequality (Methods):
\begin{equation}
    v_A \leq v_\mathrm{lim} := \sqrt{I_\mathrm{F}} := \sqrt{\braket{(\partial_t P / P)^2}},
    \label{Eq:SpeedLimitIneq}
\end{equation}
where we define the Fisher information $I_\mathrm{F}$~\cite{Frank2009,Cover2012} and the speed $v_A := |\partial_t \braket{A} - \braket{\partial_t A}| / \Delta A$, which characterizes the temporal change rate of a type-dependent quantity $A := \{ A_i \}_{i = 1}^L$ that can depend on time in general [Methods, Fig.~\ref{Fig:Overview}(a)].
Here, the average and standard deviation are defined as $\braket{A} := \sum_{i = 1}^L P_i A_i$ and $\Delta A := (\braket{A^2} - \braket{A}^2)^{1/2}$, respectively.
Inequality \eqref{Eq:SpeedLimitIneq} provides a universal upper bound on the speed of population dynamics, independent of the choice of quantity $A$ [Fig.~\ref{Fig:Overview}(b)].
We stress that \eqref{Eq:SpeedLimitIneq} applies to nonlinear dynamics though the expression is equivalent to that for Markov processes~\cite{Ito2020,Nicholson2020}, where the probability distribution follows linear dynamics.
For example, $v_\mathrm{lim}$ in \eqref{Eq:SpeedLimitIneq} can be a non-monotonic function of time, in contrast to Markovian relaxation processes, where $v_\mathrm{lim}$ decays monotonically~\cite{Ito2020}.
Note that \eqref{Eq:SpeedLimitIneq} is different from the previously obtained speed-limit inequalities in nonlinear systems~\cite{Yoshimura2021PRR,Yoshimura2021PRL}, which have been discussed mainly for chemical reaction networks.
Following Ref.~\cite{Nicholson2020}, we can interpret \eqref{Eq:SpeedLimitIneq} as the uncertainty relation between the timescale of dynamical quantities (${v_A}^{-1}$) and the information of dynamics ($\sqrt{I_\mathrm{F}}$).
\\

\begin{figure*}[t]
\includegraphics[scale=0.8]{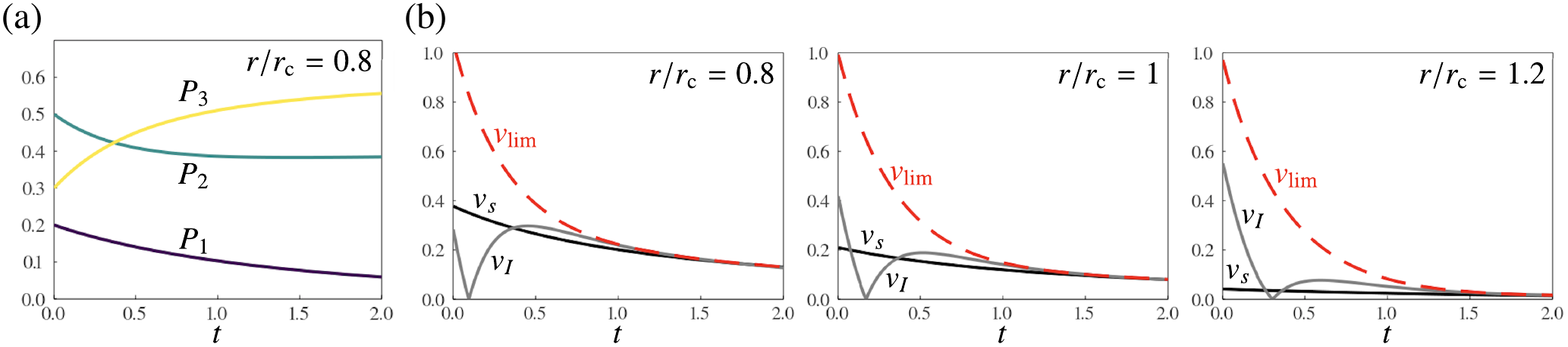}
\caption{\textbf{Speed limit for the evolutionary dynamics with natural selection and mutation.}
(a) Typical time dependence of the proportion $P_i$ for $r / r_\mathrm{c} = 0.8$.
(b) Inequality \eqref{Eq:SpeedLimitIneq} holds, regardless of the parameters ($r / r_\mathrm{c}$) and quantities (growth rate $s$ or diversity $I$).
The speed of the growth rate $v_s$ and that of change in diversity $v_I$ are compared with the speed limit $v_\mathrm{lim}$ for $r / r_\mathrm{c} = 0.8$ (left), $r / r_\mathrm{c} = 1$ (center), and $r / r_\mathrm{c} = 1.2$ (right).
See Methods for the other parameters used.}
\label{Fig:Demo}
\end{figure*}

{\noindent \textbf{Relation to Fisher's fundamental theorem.}}
Notably, our general constraint includes Fisher's fundamental theorem as a special case when applied to an evolutionary model with natural selection.
We take $F_i = s_i  N_i$ in Eq.~\eqref{Eq:GenericPopDynamics}, where $s_i > 0$ is the type-dependent growth rate.
In such systems, Fisher's fundamental theorem of natural selection asserts that the increase in the average growth rate is equal to the variance of the growth rate~\cite{Li1967}, i.e., $\partial_t \braket{s} = (\Delta s)^2$.
As shown in Methods, we find that Fisher's fundamental theorem is a special case of \eqref{Eq:SpeedLimitIneq}, $v_s = v_\mathrm{lim}$.
Note that $v_\mathrm{lim}$ in \eqref{Eq:SpeedLimitIneq} is equivalent to Crow’s index of opportunity for selection, which provides an empirical estimate of the maximum strength of natural selection acting on a given population~\cite{Crow1989,Waples2020}.
Furthermore, even when the growth rate depends on time and densities, we show that an extended version of the fundamental theorem~\cite{Edwards1994,Baez2021} is a special case of \eqref{Eq:SpeedLimitIneq} (Methods).
Our result therefore covers a variety of previous results established in population biology in light of information theory and statistical physics.
For more general dynamics with mutation, Fisher's fundamental theorem does not hold.
Nevertheless, the speed-limit inequality \eqref{Eq:SpeedLimitIneq} is satisfied and thus regarded as a generalization of the fundamental theorem.
\\

{\noindent \textbf{Speed limit for evolutionary dynamics.}}
We next consider another evolutionary model with natural selection and mutation [Fig.~\ref{Fig:Overview}(d)] by taking $F_i = s_i N_i + \sum_{j = 1}^L m_{ij} N_j$ in Eq.~\eqref{Eq:GenericPopDynamics}~\cite{Bull2005,Domingo2019}.
Here, $s_i > 0$ is the growth rate and $m_{ij} \geq 0$ ($i \neq j$) is the mutation rate from type $j$ to $i$.
To demonstrate inequality \eqref{Eq:SpeedLimitIneq}, we take $L = 3$ with $s_2 = s_3 = \bar{s}$ and examine a situation where type $1$ will survive (become extinct) after a long time if the growth rate $s_1 = \bar{s} + r$ is larger (smaller) than a critical value $\bar{s} + r_\mathrm{c}$ (Methods).
The extinction transition at $r = r_\mathrm{c}$ corresponds to the transcritical bifurcation~\cite{Strogatz2018}.

Figure~\ref{Fig:Demo}(a) shows typical time dependence of the proportion $P_i$.
As shown in Fig.~\ref{Fig:Demo}(b), regardless of the value of $r/r_\mathrm{c}$, the speed of the growth rate $v_s$ (black solid lines) is bounded by the speed limit $v_\mathrm{lim}$ (red dashed lines), which verifies  \eqref{Eq:SpeedLimitIneq}.
To confirm the generality of \eqref{Eq:SpeedLimitIneq}, we introduce the Shannon entropy $I_\mathrm{S} := \braket{I}$ with $I := \{ I_i \}_{i = 1}^L := \{ - \ln P_i \}_{i = 1}^L$~\cite{Cover2012} as the (logarithm of) diversity of population (see Supplementary Fig.~\ref{ExFig:MutSel} for typical time dependence of $I_\mathrm{S}$).
We show that the speed of change in diversity, $v_I$ (gray solid lines), is also bounded by $v_\mathrm{lim}$.
\\

{\noindent \textbf{Universal constraint around transcritical bifurcation point.}}
A notable consequence of the speed limit follows at the transcritical bifurcation point ($r=r_\mathrm{c}$), where an observable $A$ typically exhibits critical slowing down~\cite{Munoz2018,Drake2019} with a power-law decay of the speed, $v_A \sim t^{-\alpha_A}$.
While $\alpha_A$ can vary for different $A$, inequality \eqref{Eq:SpeedLimitIneq} indicates that $\alpha_A$ is bounded by a universal factor $\alpha_\mathrm{lim}$ determined by the Fisher information [Fig.~\ref{Fig:Overview}(c)].
In the evolutionary model with natural selection and mutation, we find $P_1 \sim t^{-1}$ (Methods) and thus
\begin{equation}
    v_\mathrm{lim} \sim \sqrt{(\partial_t P_1)^2 / P_1} \sim t^{- \alpha_\mathrm{lim}^\mathrm{TC}}
    \label{Eq:SpeedLimitTCB}
\end{equation}
with $\alpha_\mathrm{lim}^\mathrm{TC}=3/2$.
Then, we have
\begin{equation}
    \alpha_A \geq \alpha_\mathrm{lim}^\mathrm{TC}=3/2
    \label{Eq:ExpLimitTCB}
\end{equation}
for arbitrary $A$ in this process.

Additionally, if the parameter is slightly off the bifurcation point, the system can exhibit dynamical scaling, in a manner similar to critical phenomena~\cite{Schmittmann1995,Henkel2008,Corral2018}.
Assuming that the relaxation times of the speed and speed limit diverge at the bifurcation point as $\sim |r - r_\mathrm{c}|^{-\beta_A}$ and $\sim |r - r_\mathrm{c}|^{-\beta_\mathrm{lim}^\mathrm{TC}}$, respectively, we obtain the dynamical scaling laws as
\begin{align}
& v_A (r - r_\mathrm{c}, t) \simeq t^{-\alpha_A} f_A^\pm (t^{1 / \beta_A} |r - r_\mathrm{c}|),
\label{Eq:ScalingLaw} \\
& v_\mathrm{lim} (r - r_\mathrm{c}, t) \simeq t^{-\alpha_\mathrm{lim}^\mathrm{TC}} f_\mathrm{lim}^\pm (t^{1 / \beta_\mathrm{lim}^\mathrm{TC}} |r - r_\mathrm{c}|),
\label{Eq:ScalingLawLim}
\end{align}
where $f_A^+$ and $f_\mathrm{lim}^+$ ($f_A^-$ and $f_\mathrm{lim}^-$) are scaling functions for $r - r_\mathrm{c} >0$ ($<0$).
Combining inequality \eqref{Eq:SpeedLimitIneq} and the scaling laws \eqref{Eq:ScalingLaw} and \eqref{Eq:ScalingLawLim}, we derive another constraint on the exponents as $\beta_A \leq \beta_\mathrm{lim}^\mathrm{TC}$ (Methods).
In the numerical simulations, we have only found the case with $\beta_A = \beta_\mathrm{lim}^\mathrm{TC}$ (see below), which suggests that the diverging relaxation time of any speed should be proportional to the relaxation time of a single quantity (i.e., $P_1$ in the present model) in a similar way to critical phenomena~\cite{Schmittmann1995,Henkel2008}.

\begin{figure}[t]
\includegraphics[scale=0.8]{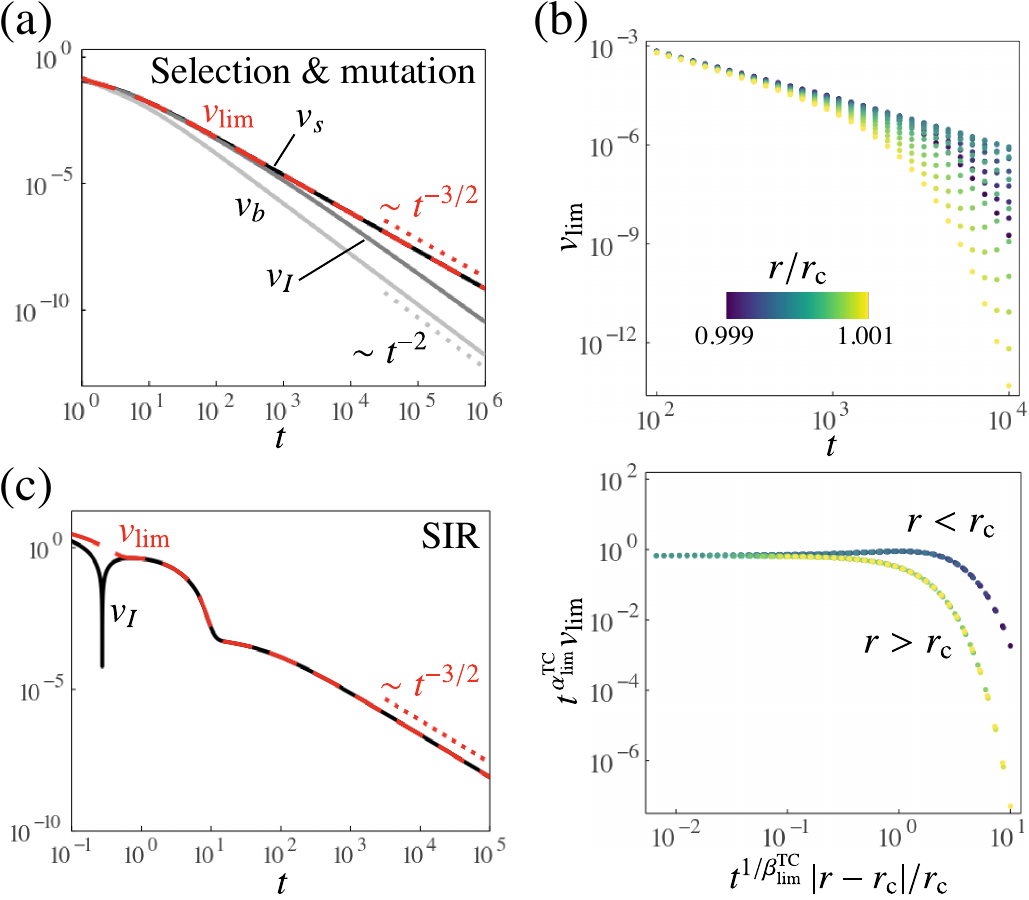}
\caption{\textbf{Universal bounds for the critical scaling exponents at the transcritical bifurcation.}
(a) Power-law decay of the speeds $v_s$, $v_I$, $v_b$, and $v_\mathrm{lim}$ at the transcritical bifurcation point ($r = r_\mathrm{c}$) of the evolutionary model with selection and mutation.
The asymptotic forms ($v_\mathrm{lim} \sim t^{-3/2}$ and $v_b \sim t^{-2}$) are shown with dotted lines.
(b) Time and parameter dependence of $v_\mathrm{lim}$ (upper panel) and the corresponding scaling plot (lower panel) near the bifurcation point ($0.999 \leq r / r_\mathrm{c} \leq 1.001$).
The exponents are given as $\alpha_\mathrm{lim}^\mathrm{TC} = 3/2$ and $\beta_\mathrm{lim}^\mathrm{TC} = 1$.
(c) Power-law decay of $v_I$ and $v_\mathrm{lim}$ at the transcritical bifurcation point of the SIR model.
The asymptotic form ($v_\mathrm{lim} \sim t^{-3/2}$) is shown with a dotted line.
For (a) and (b), we use the same parameters as those for Fig.~\ref{Fig:Demo}.
See Methods for the parameters used for (c).}
\label{Fig:TCB}
\end{figure}

To confirm the above argument, we demonstrate the long-time relaxation of $v_\mathrm{lim}$, $v_s$, $v_I$, and a speed $v_b$ for the type index $b := \{ b_i \}_{i = 1}^L := \{ i \}_{i = 1}^L$ at the bifurcation point ($r = r_\mathrm{c}$) [Fig.~\ref{Fig:TCB}(a)].
We find $v_\mathrm{lim} \sim t^{-3/2}$ [red dotted line in Fig.~\ref{Fig:TCB}(a)], which is consistent with \eqref{Eq:SpeedLimitTCB}.
We also obtain $v_s \sim t^{-3/2}$, $v_I \sim t^{-2} \ln t$, and $v_b \sim t^{-2}$ (see Methods for the derivation), and the corresponding exponents are $\alpha_s = 3 / 2$, $\alpha_I = 2$ (neglecting the logarithmic dependence), and $\alpha_b = 2$, which indeed satisfy inequality \eqref{Eq:ExpLimitTCB}.
Moreover, slightly off the bifurcation point, we find the expected scaling laws [\eqref{Eq:ScalingLaw} and \eqref{Eq:ScalingLawLim}] of $v_s$, $v_b$ (Supplementary Fig.~\ref{ExFig:ScalingTCB}), and $v_\mathrm{lim}$ [Fig.~\ref{Fig:TCB}(b)] with $\beta_s = \beta_b = \beta_\mathrm{lim}^\mathrm{TC} = 1$.

Beyond specific dynamics, we conjecture that the exponents for the power-law decay of the speeds at the bifurcation point in population dynamics are bounded by a universal constant $\alpha_\mathrm{lim}$ that only depends on the type of bifurcation.
Similarly, the exponent $\beta_\mathrm{lim}$ is also conjectured to be determined by the bifurcation type.
These conjectures are plausible because critical properties associated with the bifurcation can be essentially described by the normal form for each bifurcation type~\cite{Strogatz2018,Corral2018}.
This universal constraint on the exponents is a unique property of nonlinear dynamics, in contrast to the previous works on speed limits for linear dynamics~\cite{Ito2020,Nicholson2020}.

As a primary example, inequality \eqref{Eq:ExpLimitTCB} can be generally applied to nonlinear dynamics that undergoes an extinction transition through the transcritical bifurcation.
We consider the SIR model with birth and death [Fig.~\ref{Fig:Overview}(e)], where $N_1$, $N_2$, and $N_3$ are the densities of susceptible, infected, and recovered individuals, respectively~\cite{Kretzschmar2010} (Methods).
This model is genuinely nonlinear in that $F_i(N_1,N_2,N_3)$ in Eq.~\eqref{Eq:GenericPopDynamics} is a nonlinear function.
In this model, the transcritical bifurcation occurs as an extinction transition of the infected and recovered individuals, i.e., a transition between the disease-free and endemic states, and the critical slowing down occurs ($P_2 \sim P_3 \sim t^{-1}$) at the bifurcation point (Methods).
In Supplementary Fig.~\ref{ExFig:SIR}, we show typical time dependence of the proportion at the bifurcation point.
We find that the speed of change in diversity $v_I$ and the speed limit $v_\mathrm{lim}$ follow the same power-law decay as $v_I \sim v_\mathrm{lim} \sim t^{-3/2}$ [Fig.~\ref{Fig:TCB}(c)], satisfying the formulae \eqref{Eq:SpeedLimitTCB} and \eqref{Eq:ExpLimitTCB}.
\\

\begin{figure}[t]
\includegraphics[scale=0.8]{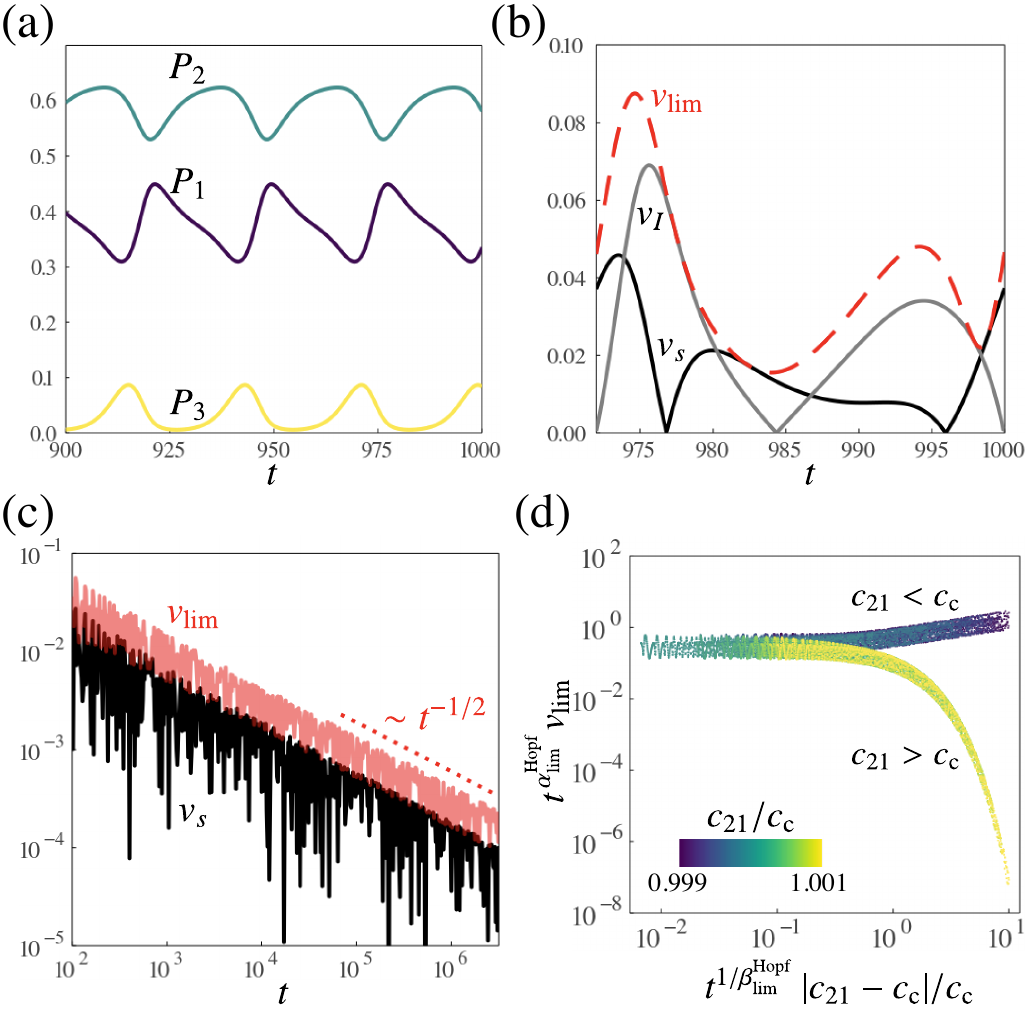}
\caption{\textbf{Universal bounds for the critical scaling exponents at the supercritical Hopf bifurcation.}
Limit-cycle oscillation of (a) the proportion $P_i$ and (b) the speeds $v_s$, $v_I$, and $v_\mathrm{lim}$ in the competitive Lotka-Volterra model.
(c) Power-law decay of $v_\mathrm{lim}$, compared with $v_s$ at the Hopf bifurcation point.
The asymptotic form of the amplitude relaxation ($v_\mathrm{lim} \sim t^{-1/2}$) is shown with a dotted line.
The curves are rattling since the number of plotted points is finite; similarly to (b), $v_s$ oscillates between zero and nonzero values, while $v_\mathrm{lim}$ stays nonzero.
(d) Scaling plot of the time and interaction dependence of $v_\mathrm{lim}$ near the bifurcation point ($0.999 \leq c_{21} / c_\mathrm{c} \leq 1.001$).
The limit cycle appears for $c_{21} < c_\mathrm{c}$, while the steady-state coexistence of three types appears for $c_{21} \geq c_\mathrm{c}$, where $c_\mathrm{c}$ is the Hopf bifurcation point (Methods).
The exponents are given as $\alpha_\mathrm{lim}^\mathrm{Hopf} = 1/2$ and $\beta_\mathrm{lim}^\mathrm{Hopf} = 1$.
See Methods for the parameters used.}
\label{Fig:SHB}
\end{figure}

{\noindent \textbf{Universal constraint around Hopf bifurcation point.}}
To verify our conjecture for other types of bifurcations, we focus on the Hopf bifurcation, at which a limit cycle starts to appear~\cite{Strogatz2018}.
According to the normal form of the supercritical Hopf bifurcation, the deviation from the steady state decays with oscillation as $\sim t^{-1/2} \cos \omega t$ at the bifurcation point (Methods).
Thus, for population dynamics undergoing the supercritical Hopf bifurcation, the proportion follows $P_i \sim \mathrm{const.} + t^{-1/2} \cos \omega t$, and the speed limit decays as
\begin{equation}
    v_\mathrm{lim} = \sqrt{\sum_{i=1}^L (\partial_t P_i)^2 / P_i} \sim t^{-\alpha_\mathrm{lim}^\mathrm{Hopf}},
    \label{Eq:SpeedLimitSHB}
\end{equation}
with $\alpha_\mathrm{lim}^\mathrm{Hopf}=1/2$, where we only consider the amplitude relaxation by neglecting the oscillatory component.
Correspondingly, if we assume a power-law decay of the speed amplitude as $v_A \sim t^{-\alpha_A}$, $\alpha_A$ should satisfy
\begin{equation}
    \alpha_A \geq \alpha_\mathrm{lim}^\mathrm{Hopf}=1/2.
    \label{Eq:ExpLimitSHB}
\end{equation}

As an ecological model that undergoes the supercritical Hopf bifurcation, we consider the competitive Lotka-Volterra model [Fig.~\ref{Fig:Overview}(f)] by taking $F_i = s_i N_i - \sum_{j = 1}^L c_{ij} N_i N_j$ in Eq.~\eqref{Eq:GenericPopDynamics}~\cite{Zeeman1993, Hofbauer1998}.
Here, $s_i$ is the growth rate, $c_{ij} > 0$ represents the competitive interaction between type $i$ and $j$, and these parameters are set around the Hopf bifurcation (Methods).

We first show typical limit-cycle oscillation of the proportion [Fig.~\ref{Fig:SHB}(a)].
Comparing $v_s$, $v_I$, and $v_\mathrm{lim}$ within a single period [Fig.~\ref{Fig:SHB}(b)], we confirm that inequality \eqref{Eq:SpeedLimitIneq} holds even when the limit cycle appears.
By tuning the parameters to the Hopf bifurcation point, we numerically find the power-law decay of the speed amplitudes~\cite{Strizhak1996} as $v_s \sim v_I \sim v_\mathrm{lim} \sim t^{-1/2}$ [Fig.~\ref{Fig:SHB}(c) and Supplementary Fig.~\ref{ExFig:LV}], verifying \eqref{Eq:SpeedLimitSHB} and \eqref{Eq:ExpLimitSHB}.
Then, changing the parameters slightly off the bifurcation point, we find that the counterparts of the scaling laws \eqref{Eq:ScalingLaw} and \eqref{Eq:ScalingLawLim} hold for the speed amplitudes [Fig.~\ref{Fig:SHB}(d) and Supplementary Fig.~\ref{ExFig:ScalingSHB}] with $\beta_s = \beta_I = \beta_\mathrm{lim}^\mathrm{Hopf} = 1$.
\\

{\noindent \textbf{Discussion}}

We have illustrated the applications of the dynamical constraint \eqref{Eq:SpeedLimitIneq} to ecological and evolutionary models.
Focusing on the bifurcation unique to nonlinear dynamics, we have argued that the exponents of speeds at critical slowing down have the universal bounds that depend only on the bifurcation type.
In particular, for the transcritical and supercritical Hopf bifurcations, we have confirmed the theoretically obtained formulae \eqref{Eq:SpeedLimitTCB}-\eqref{Eq:ExpLimitSHB} using numerical simulations.
Similar formulae are obtained for other bifurcations, e.g., $\alpha_A\geq\alpha_\mathrm{lim}^\mathrm{SN}=2$ for the saddle-node bifurcation (Methods), which appears in population dynamics~\cite{Veraart2012,Dai2012,Scheffer2015}.

Considering the probability~\cite{Ito2020,Nicholson2020} instead of the proportion, we may extend our argument to critical phenomena in many-body stochastic systems, which can express nonequilibrium phenomena different from ecological and evolutionary dynamics.
For instance, lattice gas models~\cite{Schmittmann1995}, the contact process~\cite{Henkel2008}, and biological systems such as swarms~\cite{Cavagna2017} are potentially subject to constraints corresponding to \eqref{Eq:ExpLimitTCB} or \eqref{Eq:ExpLimitSHB} with possibly irrational lower bounds.

The methodologies of ecology and evolution have been developed almost independently~\cite{Bell2017}.
However, ecological and evolutionary dynamics may not be separable in some situations.
For example, rapid evolution can occur on the same timescale as that of ecological processes when there are drastic environmental changes~\cite{Bell2017}.
General relations such as \eqref{Eq:SpeedLimitIneq} will be useful in quantitative understanding of even inseparable eco-evolutionary dynamics.
\\


{\noindent \textbf{Methods}}

{\noindent \textbf{Price equation and speed-limit inequality.}}
Purely from the conservation of the total proportion, $\sum_{i = 1}^L P_i = 1$, we can derive the Price equation~\cite{Price1972a,Frank2020},
\begin{equation}
    \partial_t \braket{A} = \mathrm{cov} (A, \partial_t P / P) + \braket{\partial_t A}.
    \label{Eq:Price}
\end{equation}
Here, $A := \{ A_i \}_{i = 1}^L$ is a generally time-dependent quantity depending on each type, e.g., the growth rate, $\braket{A} := \sum_{i = 1}^L P_i A_i$, $\mathrm{cov} (A, B) := \braket{(A - \braket{A})(B - \braket{B})}$, and $\partial_t P / P := \{ \partial_t P_i / P_i \}_{i = 1}^L$.

We define the speed of $A$ as $v_A := |\partial_t \braket{A} - \braket{\partial_t A}| / \Delta A$~\cite{Ito2020,Nicholson2020} with $\Delta A := \sqrt{\braket{(A - \braket{A})^2}}$.
The inverse of the speed, ${v_A}^{-1}$, represents the time required for $A$ to change to a statistically distinguishable value~\cite{Nicholson2020}, and thus $v_A$ characterizes the speed of the temporal change in $A$ [Fig.~\ref{Fig:Overview}(a)].
From Eq.~\eqref{Eq:Price}, we obtain the speed-limit inequality \eqref{Eq:SpeedLimitIneq} as
\begin{align}
    v_A &= \frac{|\mathrm{cov} (A, \partial_t P / P)|}{\Delta A} \nonumber \\
    &\leq \frac{1}{\Delta A} \sqrt{\sum_{i=1}^L P_i (A_i - \braket{A})^2} \sqrt{\sum_{i=1}^L P_i \left( \frac{\partial_t P_i}{P_i} \right)^2} \nonumber \\
    &= \sqrt{\braket{\left( {\partial_t P} / P \right)^2}}, \nonumber
\end{align}
where $\braket{\partial_t P / P} = \partial_t \sum_{i=1}^L P_i = 0$ and the Cauchy-Schwarz inequality are used in the second line.

The equal sign in \eqref{Eq:SpeedLimitIneq} is achieved when $A - \braket{A}$ is parallel or anti-parallel to $\partial_t P / P$ in the $L$-dimensional vector space, where $\partial_t P_i / P_i$ is regarded as the $i$th component of $\partial_t P / P$, for instance.
When there are only two types ($L = 2$), we can explicitly obtain $(A_1 - \braket{A}, A_2 - \braket{A}) = (A_1 - A_2) (P_2, -P_1)$ and $(\partial_t P_1 / P_1, \partial_t P_2 / P_2) = [\partial_t P_1 / (P_1 P_2)] (P_2, -P_1)$, leading to $A - \braket{A} \parallel \partial_t P / P$, and the equal sign in \eqref{Eq:SpeedLimitIneq} is achieved regardless of the details of dynamics.
\\

{\noindent \textbf{Fisher's fundamental theorem of natural selection.}}
We first take $F_i = s_i  N_i$ in Eq.~\eqref{Eq:GenericPopDynamics}, where $s_i > 0$ is the type-dependent growth rate.
Since $\partial_t P_i = (s_i - \braket{s}) P_i$ from Eq.~\eqref{Eq:GenericFreqDynamics}, Eq.~\eqref{Eq:Price} leads to
\begin{equation}
    \partial_t \braket{s} = \mathrm{cov} (s, s - \braket{s}) = (\Delta s)^2,
    \label{Eq:Price_Evo}
\end{equation}
which means that the rate of increase in the average growth rate is equal to the variance of the growth rate, known as Fisher's fundamental theorem of natural selection~\cite{Li1967,Price1972b,Frank1992,Frank2009,Grafen2015AMNAT,Grafen2015JTB,Grafen2018}.
On the other hand, the equation $\partial_t P_i = (s_i - \braket{s}) P_i$ also leads to $v_\mathrm{lim} = [\braket{(s - \braket{s})^2}]^{1/2} = \Delta s$.
Since $v_s = |\partial_t \braket{s}| / \Delta s$ by definition, Eq.~\eqref{Eq:Price_Evo} is equivalent to $v_s = v_\mathrm{lim}$, suggesting that Fisher's fundamental theorem is nothing but a special case of \eqref{Eq:SpeedLimitIneq}.

We next take $F_i = f_i (\{ N_i \}_{i=1}^L, t) N_i$ in Eq.~\eqref{Eq:GenericPopDynamics} with an arbitrary function $f_i$, which is the so-called fitness~\cite{Baez2021} and represents the growth rate that can depend on the effects of interactions among types.
From Eqs.~\eqref{Eq:GenericFreqDynamics} and \eqref{Eq:Price}, we obtain $\partial_t \braket{f} - \braket{\partial_t f} = (\Delta f)^2$, which is known as an extended version of Fisher's fundamental theorem~\cite{Edwards1994,Baez2021}.
Since we can obtain $v_\mathrm{lim} = \Delta f$ and $v_f = |\partial_t \braket{f} - \braket{\partial_t f}| / \Delta f$ in the same way as explained above, the extended version of the fundamental theorem is equivalent to a special case of \eqref{Eq:SpeedLimitIneq}, $v_f = v_\mathrm{lim}$.
\\

{\noindent \textbf{Evolutionary model with natural selection and mutation.}}
We take $F_i = s_i N_i + \sum_{j = 1}^L m_{ij} N_j$ in Eq.~\eqref{Eq:GenericPopDynamics}~\cite{Bull2005,Domingo2019}, where $s_i > 0$ is the growth rate, $m_{ij} \geq 0$ ($i \neq j$) is the mutation rate from type $j$ to $i$, and $\sum_{i = 1}^L m_{ij} = 0$ for all $j$.
Note that we here consider a large and well-mixed population where noise and spatial effects~\cite{Lavrentovich2013,Lavrentovich2016} are negligible.
In the following, we further assume $s_i = \bar{s} + r \delta_{i1}$ ($\bar{s}, r > 0$), $m_{1j} = 0$ for $j \neq 1$, and $m_{ij} > 0$ for $i \neq 1$ and $i \neq j$, where $\delta_{ij}$ is the Kronecker delta.
From Eq.~\eqref{Eq:GenericFreqDynamics}, we can obtain the equation for $P_1$ as $\partial_t P_1 = (r + m_{11} - r P_1) P_1$.
Thus, $r_\mathrm{c}$ ($:= -m_{11} = \sum_{i = 2}^L m_{i1}$) is a transcritical bifurcation point~\cite{Strogatz2018}: $P_1^\mathrm{s} = 0$ for $r \leq r_\mathrm{c}$, while $P_1^\mathrm{s} = (r - r_\mathrm{c}) / r > 0$ for $r > r_\mathrm{c}$, provided $P_1 (t = 0) > 0$, where $P_i^\mathrm{s} := P_i (t \to \infty)$ is the steady-state proportion of type $i$.
The qualitative change in $P_1^\mathrm{s}$ at $r = r_\mathrm{c}$ represents the transition between survival and extinction of type $1$.

At the bifurcation point ($r = r_\mathrm{c}$), $P_1$ satisfies
\begin{equation}
    \partial_t P_1 = - r_\mathrm{c} {P_1}^2,
    \label{Eq:P1eq_TCB}
\end{equation}
which leads to $P_1 \sim t^{-1}$ after a long time.
If a quantity $A$ is independent of time, e.g., $A = s$ (growth rate) or $A = b$ (type index), we can obtain $\partial_t \braket{A} \sim t^{-2}$ according to $P_1 \sim t^{-1}$.
Then, assuming that $\Delta A$ shows a power-law decay as $\Delta A \sim t^{-\delta_A}$ with a certain exponent $\delta_A$, we obtain $v_A = |\partial_t \braket{A}| / \Delta A \sim t^{-2 + \delta_A}$.
On the other hand, if $A$ is time-dependent, e.g., $A = I = - \ln P$ (diversity), the asymptotic time dependence of $\partial_t \braket{A} - \braket{\partial_t A}$, $\Delta A$, and $v_A = |\partial_t \braket{A} - \braket{\partial_t A}| / \Delta A$ generally depends on the time dependence of $A$.
Thus, the asymptotic forms of the speeds follow $v_s \sim t^{-3/2}$ since $\Delta s = [P_1 (1 - P_1)]^{1/2} r_\mathrm{c} \sim t^{-1/2}$ (i.e., $\delta_s = 1/2$), $v_b \sim t^{-2}$ since $\Delta b (t \to \infty) \neq 0$ (i.e., $\delta_b = 0$), and $v_I \sim t^{-2} \ln t$ since $|\partial_t \braket{I} - \braket{\partial_t I}| \sim t^{-2} \ln t$ and $\Delta I (t \to \infty) \neq 0$.

Near but off the bifurcation point ($r \simeq r_\mathrm{c}$), we can linearize the equation of $P_1$ after a long time as
\begin{equation}
    \partial_t (P_1 - P_1^\mathrm{s}) \simeq - |r - r_\mathrm{c}| (P_1 - P_1^\mathrm{s}),
\end{equation}
both for $r > r_\mathrm{c}$ and $r < r_\mathrm{c}$.
Thus, $P_1$ shows an exponential relaxation with the relaxation time proportional to $|r - r_\mathrm{c}|^{-1}$, which diverges at the bifurcation point.
Such divergence suggests that the relaxation times $\tau_A$ (for $v_A$) and $\tau_\mathrm{lim}$ (for $v_\mathrm{lim}$) should also diverge at the bifurcation point as $\tau_A \sim |r - r_\mathrm{c}|^{-\beta_A}$ and $\tau_\mathrm{lim} \sim |r - r_\mathrm{c}|^{-\beta_\mathrm{lim}^\mathrm{TC}}$ with certain exponents $\beta_A$ and $\beta_\mathrm{lim}^\mathrm{TC}$.
This motivates us to consider the dynamical scaling laws \eqref{Eq:ScalingLaw} and \eqref{Eq:ScalingLawLim} since similar dynamical scaling is applied to order parameters with diverging relaxation time in critical phenomena~\cite{Schmittmann1995,Henkel2008}.
Since inequality \eqref{Eq:SpeedLimitIneq} indicates that $v_A$ should show an exponential decay earlier than $v_\mathrm{lim}$, we obtain an inequality between the relaxation timescales as $\tau_A \leq \tau_\mathrm{lim}$ near the bifurcation point, which leads to $\beta_A \leq \beta_\mathrm{lim}^\mathrm{TC}$.

For Figs.~\ref{Fig:Demo}, \ref{Fig:TCB}(a), \ref{Fig:TCB}(b), Supplementary Figs.~\ref{ExFig:MutSel}, and \ref{ExFig:ScalingTCB}, we used the mutation rates
\begin{equation}
    (m_{ij}) =
    \begin{pmatrix}
        -2.1 & 0 & 0 \\
        1 & -1.2 & 0.8 \\
        1.1 & 1.2 & -0.8
    \end{pmatrix}  \nonumber
\end{equation}
with $r_\mathrm{c} = -m_{11} = 2.1$ and the initial state $(P_1, P_2, P_3)|_{t = 0} = (0.2, 0.5, 0.3)$.
\\

{\noindent \textbf{SIR model with birth and death.}}
We take $L = 3$, $F_1 = 1 - \lambda N_1 N_2 / N_\mathrm{tot} - N_1$, $F_2 = \lambda N_1 N_2 / N_\mathrm{tot} - \gamma N_2 - N_2$, and $F_3 = \gamma N_2 - N_3$ in Eq.~\eqref{Eq:GenericPopDynamics}, where $N_1$, $N_2$, and $N_3$ are the densities of susceptible, infected, and recovered individuals, respectively~\cite{Kretzschmar2010}.
Here, $\lambda$ is the infection rate, $\gamma$ is the recovery rate, and we take both the birth and death rates as unity by rescaling time $t$ and $N_\mathrm{tot}$, where the death rate is assumed to be the same for all three types.
Assuming a nonzero proportion of the infected individuals at the initial time [$P_2 (t = 0) > 0$], we can obtain the steady-state proportions as
\begin{equation}
    (P_1^\mathrm{s}, P_2^\mathrm{s}, P_3^\mathrm{s}) =
    \left\{
    \begin{array}{lr}
        (1, 0, 0) & (\lambda \leq \lambda_\mathrm{c}) \\
        \left( \frac{\lambda_\mathrm{c}}{\lambda}, \frac{\lambda - \lambda_\mathrm{c}}{\lambda_\mathrm{c} \lambda}, \frac{(\lambda_\mathrm{c} - 1) (\lambda - \lambda_\mathrm{c})}{\lambda_\mathrm{c} \lambda} \right)  & (\lambda > \lambda_\mathrm{c})
    \end{array}, \nonumber
    \right.
\end{equation}
where the extinction transition for the infected and recovered individuals occurs through the transcritical bifurcation at $\lambda = \lambda_\mathrm{c}$ $(:= \gamma + 1)$.

We consider the long-time dynamics of $N_i$ at the bifurcation point ($\lambda = \lambda_\mathrm{c}$).
First, since $N_\mathrm{tot}$ shows an exponential relaxation according to $\partial_t N_\mathrm{tot} = 1 - N_\mathrm{tot}$, $N_\mathrm{tot} \simeq 1$ after a long time.
Defining the deviation from the steady state as $\Delta N_i := N_i - N_i^\mathrm{s}$ with $(N_1^\mathrm{s}, N_2^\mathrm{s}, N_3^\mathrm{s}) := (1, 0, 0)$, we can linearize Eq.~\eqref{Eq:GenericPopDynamics} as $\partial_t \Delta N_1 = - \Delta N_1 - \lambda_\mathrm{c} \Delta N_2$, $\partial_t \Delta N_2 = 0$, and $\partial_t \Delta N_3 = - \Delta N_3 + (\lambda_\mathrm{c} - 1) \Delta N_2$.
These linearized equations suggest that $N_1$ and $N_3$ should adiabatically follow the dynamics of $N_2$, which is expected to show a power-law decay if nonlinearity is taken into account.
Thus, we apply the adiabatic approximation ($\partial_t N_1 \simeq 0$) to the equation for $N_1$ and obtain $N_1 \simeq (1 + \lambda_\mathrm{c} N_2 / N_\mathrm{tot})^{-1} \simeq (1 + \lambda_\mathrm{c} N_2)^{-1}$ on the timescale where $N_2$ changes.
Then, $\partial_t N_2 = \lambda_\mathrm{c} (N_1 / N_\mathrm{tot} - 1) N_2 \simeq - {\lambda_\mathrm{c}}^2 {N_2}^2 + O({N_3}^3)$, leading to a power-law decay of $N_2$ as expected: $N_2 \simeq ({\lambda_\mathrm{c}}^2 t)^{-1}$.
Lastly, applying the adiabatic approximation ($\partial_t N_3 \simeq 0$) to the equation for $N_3$, we obtain $N_3 \simeq (\lambda_\mathrm{c} - 1) N_2 \simeq (\lambda_\mathrm{c} - 1) ({\lambda_\mathrm{c}}^2 t)^{-1}$.
In terms of the proportion, $P_2 \sim P_3 \sim t^{-1}$ is followed.
Regarding the power-law decay of the speed of change in diversity, we obtain $v_I \sim t^{-3/2}$ since $|\partial_t \braket{I} - \braket{\partial_t I}| = |\sum_i (\partial_t P_i) \ln P_i| \sim t^{-2} \ln t$ and $\Delta I = [\sum_i P_i (\ln P_i)^2 - (\sum_i P_i \ln P_i)^2]^{1/2} \sim t^{-1/2} \ln t$.

For Fig.~\ref{Fig:TCB}(c) and Supplementary Fig.~\ref{ExFig:SIR}, we used the recovery rate $\gamma = 15$, the infection rate $\lambda = \lambda_\mathrm{c} = 16$, and the initial state $(N_1, N_2, N_3)|_{t = 0} = (1.1, 0.1, 0)$.
\\

{\noindent \textbf{Power-law decay at the supercritical Hopf bifurcation.}}
The normal form of the supercritical Hopf bifurcation is given as $\partial_t z = (\mu + \mathrm{i} \omega) z - |z|^2 z$, where $z$ is a complex variable and $\omega > 0$~\cite{Strogatz2018}.
The only fixed point is $z = 0$ for $\mu \leq 0$, while the limit cycle appears with the amplitude $|z| = \sqrt{\mu}$ and the period $2 \pi / \omega$ for $\mu > 0$.
At the bifurcation point ($\mu = 0$), the amplitude follows $\partial_t |z| = -|z|^3$, which leads to a power-law decay of the amplitude as $|z| \sim t^{-1/2}$ and correspondingly an oscillatory decay of $\mathrm{Re} \, z$ or $\mathrm{Im} \, z$ as $\sim t^{-1/2} \cos \omega t$.
\\

{\noindent \textbf{Competitive Lotka-Volterra model.}}
We take $F_i = s_i N_i - \sum_{j = 1}^L c_{ij} N_i N_j$ in Eq.~\eqref{Eq:GenericPopDynamics}, where $s_i$ is the growth rate, and $c_{ij} > 0$ represents the competitive interaction between type $i$ and $j$~\cite{Hofbauer1998}.
Using the previously obtained bifurcation diagram~\cite{Mohd2019} as a reference, we take $L= 3$, $(s_1, s_2, s_3) = (19, 3/2, 12)$, and
\begin{equation}
    (c_{ij}) =
    \begin{pmatrix}
        1 & 2 & 4 \\
        c_{21} & 1/6 & 1/3 \\
        1 & 1 & 3
    \end{pmatrix}. \nonumber
\end{equation}
Within a certain range of $c_{21}$ and initial states, the limit cycle appears for $c_{21} < c_\mathrm{c}$, while the steady-state coexistence of three types appears for $c_{21} \geq c_\mathrm{c}$, where $c_\mathrm{c}$ is the supercritical Hopf bifurcation point, and the numerically found value is $c_\mathrm{c} = 0.064163908$.

We used $c_{21} = 0.063833$ for Figs.~\ref{Fig:SHB}(a) and (b), $c_{21} = c_\mathrm{c}$ for Fig.~\ref{Fig:SHB}(c) and Supplementary Fig.~\ref{ExFig:LV}, and $0.999 c_\mathrm{c} \leq c \leq 1.001 c_\mathrm{c}$ for Fig.~\ref{Fig:SHB}(d) and Supplementary Fig.~\ref{ExFig:ScalingSHB}, with the initial state $(N_1, N_2, N_3)|_{t = 0} = (20, 5, 2)$.
To solve the differential equations \eqref{Eq:GenericPopDynamics}, we used a Julia package DifferentialEquations.jl~\cite{Rackauckas2017}.
\\

{\noindent \textbf{Power-law decay at the saddle-node bifurcation.}}
The normal form of the saddle-node bifurcation is given as $\partial_t x = \mu - x^2$, where the stable fixed point ($x = \sqrt{\mu}$) appears only for $\mu > 0$~\cite{Strogatz2018}.
At the bifurcation point ($\mu = 0$), we can obtain a power-law decay as $x \sim t^{-1}$.

Considering a population dynamics that undergoes the saddle-node bifurcation as an abrupt change in the density and proportion of type $1$, we should obtain $P_1 \sim \mathrm{const.} + t^{-1}$ at the bifurcation point for a certain range of initial states.
Then, the speed limit will follow $v_\mathrm{lim} = [\sum_i (\partial_i P_i)^2 / P_i]^{1/2} \sim t^{-\alpha_\mathrm{lim}^\mathrm{SN}}$ with $\alpha_\mathrm{lim}^\mathrm{SN} = 2$.
\\





\input{output.bbl}

\vspace{10pt}
{\noindent \textbf{Acknowledgements}}

We thank the Information Theory Study Group and Biology Seminar members in RIKEN iTHEMS and Kyogo Kawaguchi for scientific discussions.
We also thank Takashi Okada, Takaki Yamamoto, and Yohsuke T. Fukai for helpful comments.
This work was supported by JSPS KAKENHI Grant Numbers JP20K14435 (to K.A.), JP19K22457, JP19K23768, JP20K15882 (to R.I.), and RIKEN iTHEMS.
\\

{\noindent \textbf{Author contributions}}

K.A. and R.I. conceived the project.
K.A., R.I., and R.H. performed the analytic calculations.
K.A. performed the simulations and made all the plots.
K.A. drafted the initial version of the manuscript.
K.A., R.I., and R.H. discussed the results and wrote the manuscript.
\\




\clearpage
\onecolumngrid
\setcounter{figure}{0}
\setcounter{page}{1}
\renewcommand{\figurename}{Supplementary Fig.}

\begin{center} 
	\textbf{\Large Supplementary information} \\
	\vspace{1mm}
	\textbf{\Large Universal constraint on nonlinear population dynamics} \\
	\vspace{5mm}
	Kyosuke Adachi,$^{1,2}$ Ryosuke Iritani,$^{2,3}$ and Ryusuke Hamazaki$^{4,2}$\\
	\vspace{2mm}
	$^1$\textit{Nonequilibrium Physics of Living Matter RIKEN Hakubi Research Team,}\\
    \textit{RIKEN Center for Biosystems Dynamics Research (BDR),}\\
    \textit{2-2-3 Minatojima-minamimachi, Chuo-ku, Kobe 650-0047, Japan}\\
    $^2$\textit{RIKEN Interdisciplinary Theoretical and Mathematical Sciences Program (iTHEMS), 2-1 Hirosawa, Wako 351-0198, Japan}
    $^3$\textit{Department of Biological Sciences, Graduate School of Science,}\\
    \textit{University of Tokyo, 7-3-1 Hongo, Bunkyo-ku, Tokyo 113-0033, Japan}\\
    $^4$\textit{Nonequilibrium Quantum Statistical Mechanics RIKEN Hakubi Research Team,}\\
    \textit{RIKEN Cluster for Pioneering Research (CPR), 2-1 Hirosawa, Wako 351-0198, Japan}
\end{center}

\begin{figure*}[ht]
\includegraphics[scale=0.8]{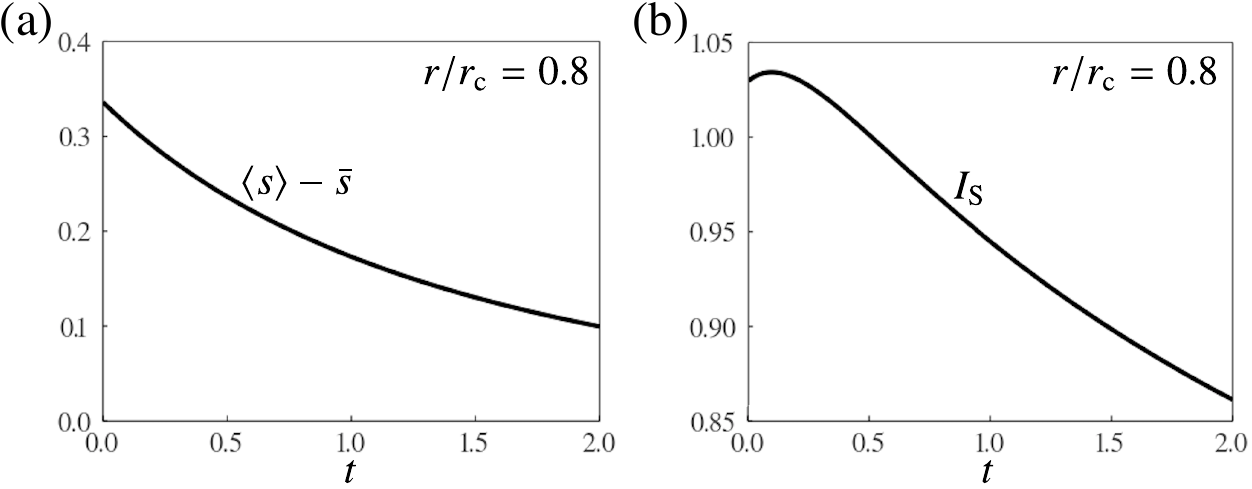}
\caption{\textbf{Typical time dependence of averaged quantities in the evolutionary model with natural selection and mutation.}
Time dependence of (a) the average growth rate $\braket{s}$ and (b) the Shannon entropy $I_\mathrm{S}$ for the same parameters used in Fig.~\ref{Fig:Demo}(a) and the left panel of Fig.~\ref{Fig:Demo}(b).}
\label{ExFig:MutSel}
\end{figure*}

\begin{figure*}[ht]
\includegraphics[scale=0.8]{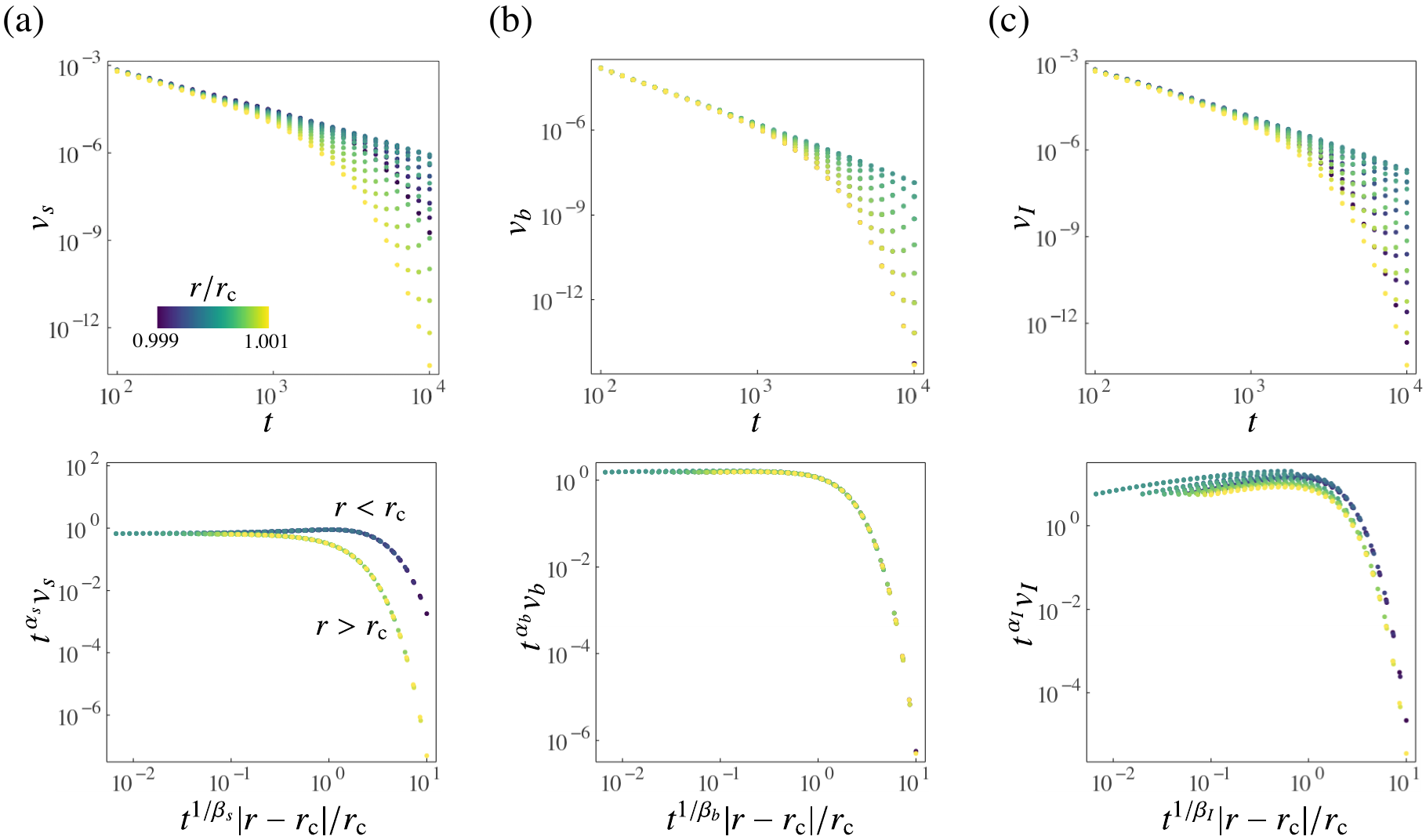}
\caption{\textbf{Dynamical scaling for the speeds of different quantities at the transcritical bifurcation.}
Time and parameter dependence of the speeds (a) $v_s$, (b) $v_b$, and (c) $v_I$ around the transcritical bifurcation point ($0.999 \leq r / r_\mathrm{c} \leq 1.001$) of the evolutionary model with selection and mutation (upper panels).
The corresponding scaling plots are shown with $\alpha_s = 3/2$, $\alpha_b = \alpha_I = 2$, and $\beta_s = \beta_b = \beta_I = 1$ (lower panels).
Note that $v_b (r - r_\mathrm{c}, t)$ seems to be almost independent of the sign of $r - r_\mathrm{c}$ in the shown parameter regime.
Also, $v_I (r - r_\mathrm{c}, t)$ does not satisfy the scaling law due to the logarithmic time dependence at $r = r_\mathrm{c}$ (i.e., $v_I \sim t^{-2} \ln t$).
For all figures, we used the same parameters as those for Fig.~\ref{Fig:TCB}(b).}
\label{ExFig:ScalingTCB}
\end{figure*}

\begin{figure*}[ht]
\includegraphics[scale=0.8]{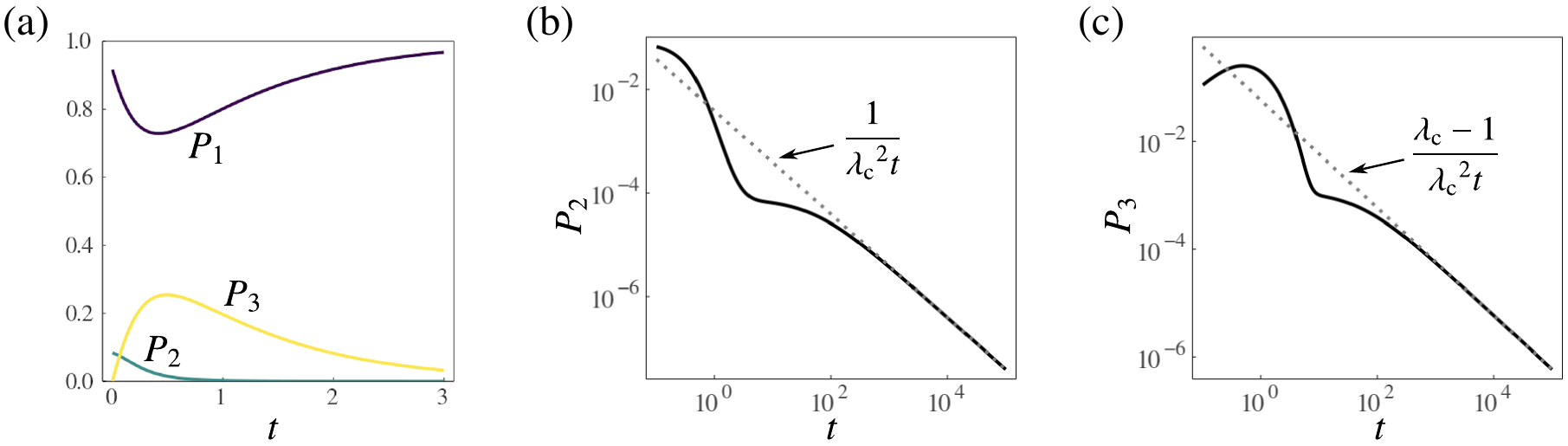}
\caption{\textbf{Typical time dependence of the proportion at the transcritical bifurcation point of the SIR model.}
(a) Time dependence of the proportion in a short timescale.
Long-time decay of the proportions of (b) the infected individuals and (c) the recovered individuals.
In (b) and (c), the asymptotic forms [$P_2 \simeq ({\lambda_\mathrm{c}}^2 t)^{-1}$ and $P_3 \simeq (\lambda_\mathrm{c} - 1) ({\lambda_\mathrm{c}}^2 t)^{-1}$] are shown with dotted lines (see Methods for the derivation).
For all figures, we used the same parameters as those for Fig.~\ref{Fig:TCB}(c).}
\label{ExFig:SIR}
\end{figure*}

\begin{figure*}[ht]
\includegraphics[scale=0.8]{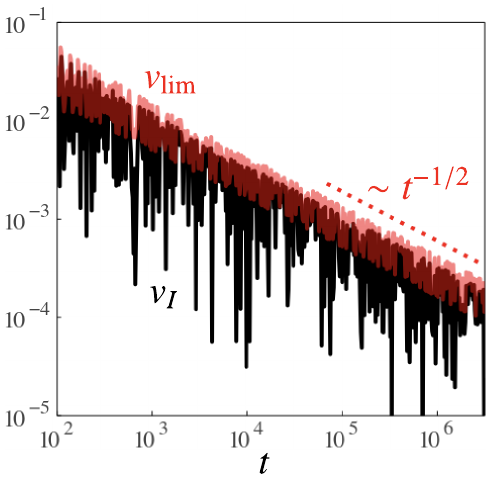}
\caption{\textbf{Power-law decay of the speed of change in diversity at the supercritical Hopf bifurcation.}
We plot the time dependence of $v_I$ and $v_\mathrm{lim}$ at the Hopf bifurcation point of the competitive Lotka-Volterra model.
The asymptotic form ($v_\mathrm{lim} \sim t^{-1/2}$) is shown with a dotted line.
We used the same parameters as those for Fig.~\ref{Fig:SHB}(c).}
\label{ExFig:LV}
\end{figure*}

\begin{figure*}[ht]
\includegraphics[scale=0.8]{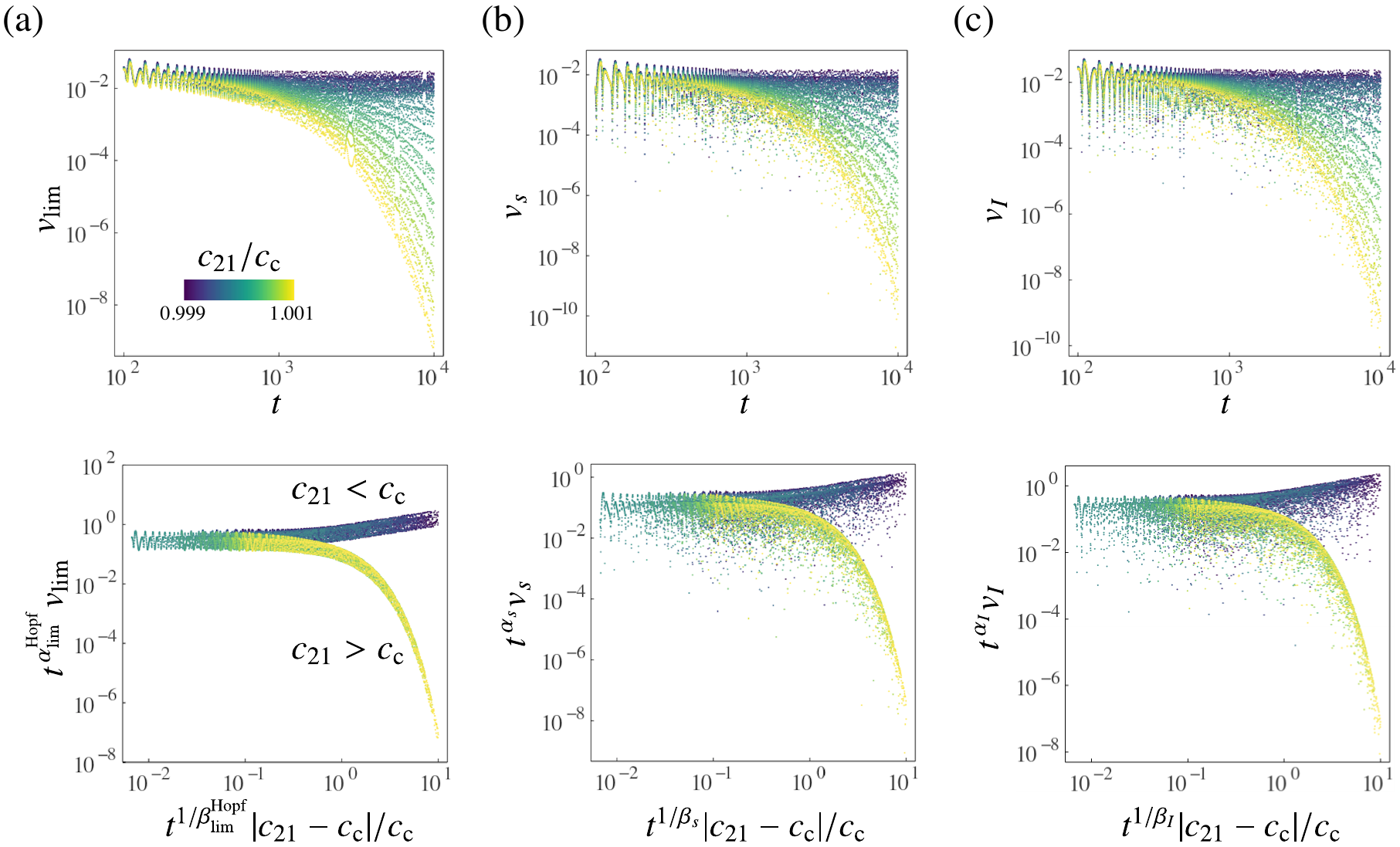}
\caption{\textbf{Dynamical scaling for speeds of different quantities at the supercritical Hopf bifurcation.}
Time and interaction dependence of (a) the speed limit $v_\mathrm{lim}$ and the speeds (b) $v_s$ and (c) $v_I$ around the Hopf bifurcation point of the competitive Lotka-Volterra model (upper panels).
The corresponding scaling plots are shown with $\alpha_\mathrm{lim}^\mathrm{Hopf} = \alpha_s = \alpha_I = 1/2$ and $\beta_\mathrm{lim}^\mathrm{Hopf} = \beta_s = \beta_I = 1$ (lower panels).
In the lower panel of (a), the same data as plotted in Fig.~\ref{Fig:SHB}(d) are reproduced for completeness.
The amplitudes of oscillating $v_\mathrm{lim}$, $v_s$, and $v_I$ follow the scaling laws.
See Methods for the parameters used.}
\label{ExFig:ScalingSHB}
\end{figure*}

\end{document}

%% file: output.bbl
%